\documentclass[proc_ichep08_slac_one]{revtex4} 
\usepackage{graphicx}
\usepackage{fancyhdr}
\usepackage{latexsym,amssymb,amsmath,amsfonts,epsfig,graphics,xspace} 
\newcommand{\DS}[1]{$\mathsf{#1}$\xspace} 

\newcommand{\GammaVM}{\Gamma_\text{VM}}         
\newcommand{\gammaV}{\gamma}                    

\newcommand{\bdi}{\begin{displaymath}}
\newcommand{\edi}{\end{displaymath}}
\newcommand{\bfi}{\begin{figure}}
\newcommand{\efi}{\end{figure}}

\newcommand{\beq}{\begin{equation}}
\newcommand{\eeq}{\end{equation}}
\newcommand{\beqa}{\begin{eqnarray}}
\newcommand{\eeqa}{\end{eqnarray}}

\newcommand{\dd}{\mathrm{d}}                    
\begin{document}

\title{Lorentz Invariance, Vacuum Energy, and Cosmology}

\author{F.R. Klinkhamer}
\affiliation{Institute for Theoretical Physics, University of Karlsruhe (TH),
76128 Karlsruhe, Germany}

\begin{abstract}
This contribution reviews recent work on a new approach to the cosmological
constant problem, which starts from the macroscopic behavior of a conserved
relativistic  microscopic variable $q$. First, the statics of the vacuum energy
density is discussed and, then, the dynamics in a cosmological context.
\end{abstract}

\maketitle

\thispagestyle{fancy}

\setcounter{section}{-1}  
\section{INTRODUCTION}
\label{sec:introduction}

The cosmological constant $\Lambda$ was introduced
by Einstein~\cite{Einstein1917} nearly a century ago and, in the years after, 
was interpreted as a possibly dynamic vacuum energy density
$\rho_\text{V}$ by Bronstein, Lema\^{i}tre,
and others~\cite{Bronstein1933-etal}. The task, then, was to
explain the apparently zero value of $\Lambda$.

But, thanks to progress in observational cosmology over the last decennium,
there are now really three cosmological constant problems
(see, e.g., Ref.~\cite{Padmanabhan2003} for one review article out of many):
\begin{enumerate}
\item
why is $|\rho_\text{V}| \ll       \big(E_\text{Planck}\big)^4
                        \approx   \big(10^{28}\:\text{eV}\big)^4 \;$?
\item
why is $\rho_\text{V} \ne 0\;$?
\item
why is $\rho_\text{V} \sim    \rho_\text{M,\,present}
                      \approx \big(10^{-3}\:\text{eV}\big)^4\;$?
\end{enumerate}
Clearly, we first need to get a handle on the main cosmological constant
problem No. 1.
Only then can we start worrying about the more delicate problems Nos. 2 and 3.

This contribution gives a brief overview of ongoing work with
G.E. Volovik~\cite{KlinkhamerVolovik2008-PRD77,KlinkhamerVolovik2008-PRD78,
KlinkhamerVolovik2008-JETPL,Klinkhamer2008}.
As the title makes clear, the write-up consists of
three parts and, correspondingly, has three conclusions
(with the last one split up into Conclusions 3.1 and 3.2).
Natural units with $c=\hbar=1$ are used, except when stated otherwise.

\section{LORENTZ INVARIANCE}
\label{sec:lorentz-invariance}

Lorentz invariance (LI) has been tested by many experiments over the years
and verified to ever greater precision (and, hopefully, accuracy).
A particularly clean and compelling case occurs with the search of
Lorentz-violating (LV) effects in the photonic sector.
\emph{A priori,} there can be $\text{O}(1)$
effects in the modified-Maxwell action, which maintain gauge
invariance, \DS{CPT}, and renormalizability. These effects are
characterized by 19 real dimensionless parameters and LI holds
only if all parameters vanish.

 From a variety of methods, it is found that the absolute value of
\emph{each} of the 19 parameters is less than $10^{-15}$, with some
even bounded down to the $10^{-32}$ level.
This result would seem to suggest that all parameters would be
strictly zero, so that local LI would be exact.
The fundamental underlying principle would remain to be discovered.
For \mbox{further discussion and references, we refer to a recent
review article\,[8(a)]                       
and an update in these Proceedings\,[8(b)].} 

\textbf{Conclusion 1:} \emph{Lorentz invariance of the electromagnetic sector
in particular has been verified to high precision.}

\section{VACUUM ENERGY}
\label{sec:vacuum-energy}

As mentioned in the Introduction, observational
cosmology suggests a nonzero value of the cosmological constant, $\Lambda>0$,
or gravitating vacuum energy density, $\rho_\text{V}=-P_\text{V}>0$.
But what is the theory behind this result?

First, consider the \emph{statics} of ``dark energy,''
viewed as vacuum energy. A simple picture~\cite{KlinkhamerVolovik2008-PRD77}
has been suggested which is based on the assumption that
the perfect quantum vacuum is:
\begin{enumerate}
\item
a \textbf{Lorentz-invariant state}
(cf. Sec.~\ref{sec:lorentz-invariance});
\item
a \textbf{self-sustained medium} at zero external pressure
(isolated from the environment);
\item
characterized by a \textbf{new type of conserved charge density}
$\boldsymbol{q}$, which is constant over spacetime.
\end{enumerate}
The analog of $q$ appears in condensed-matter physics for the so-called
Larkin--Pikin effect (1969) in magnetic phase transitions of crystals.
There, the macroscopic elastic mode with 3--momentum $\vec{k}=0$ plays a
crucial role: the coupling of this mode to the order parameter transforms
the second-order phase transition to a first-order one.

Continuing with our simple picture of the perfect quantum vacuum,
the combination of thermodynamics (Gibbs free energy),
charge conservation ($Q\equiv q\, V=\text{const}$),
and pressure equilibrium ($P_\text{V}=P_\text{ext}$)
gives an integrated form of the Gibbs--Duhem equation
(in this context, first discussed by Volovik~\cite{Volovik2003}):
\beq\label{eq:Gibbs-Duhem}
\widetilde\epsilon_\text{V}(q_0) \equiv
\bigg[\epsilon(q) - q\:\frac{\dd \epsilon(q)}{\dd q}\;\bigg]_{q=q_0}
=-P_\text{V}(q_0)=-P_\text{ext}=0~,
\eeq
for the self-tuned equilibrium value $q_0$ of the vacuum variable $q$ and
the microscopic energy density $\epsilon(q)$ from the action [the quantity
$\dd \epsilon/\dd q$ plays the role of a chemical potential $\mu$].
The following remarks are in order:
\begin{itemize}
\item
the effective energy density
$\widetilde\epsilon_\text{V}(q_0)$ is zero by cancelation of two terms which
can each be of order $\big(E_\text{Planck}\big)^4$;
\item
it can be argued that the effective energy density
$\widetilde\epsilon_\text{V}$ gravitates and $\epsilon$ not,
so that $\rho_\text{V} = \widetilde\epsilon_\text{V}(q_0)$;
\item
explicit dynamical models are known~\cite{KlinkhamerVolovik2008-PRD77},
which give indeed the result $\rho_\text{V}= \widetilde\epsilon_\text{V}(q_0)$.
\end{itemize}
All in all, we have for a perfect LI quantum vacuum
in equilibrium~\cite{KlinkhamerVolovik2008-PRD77}
\beq
\Lambda \;\equiv\; \rho_\text{V} \;=\; \widetilde{\epsilon}_\text{V}
\;\stackrel{\text{\textcircled{\sffamily\tiny 1}}}{=}\; -P_\text{V}
\;\stackrel{\text{\textcircled{\sffamily\tiny 2}}}{=}\; -P_\text{ext}
\;=\;0~,
\label{eq:Lambda-zero}
\eeq
with
step \textcircled{\sffamily\tiny 1}  from thermodynamics and
step \textcircled{\sffamily\tiny 2} from pressure equilibrium.
Result \eqref{eq:Lambda-zero} points toward a possible solution of the
cosmological constant problem No. 1, as mentioned in the Introduction.

Up till now, we considered the statics of a perfectly Lorentz-invariant
vacuum state. But what happens if the ground state itself is Lorentz
noninvariant (the physical laws are assumed to be exactly Lorentz invariant).
For example, in the presence of thermal matter (i.e., a Lorentz-noninvariant state),
pressure equilibrium gives
\beq
P_\text{V}+P_\text{M}=P_\text{ext}=0\,.
\eeq
With the resulting relation $P_\text{V}=-P_\text{M}$,
the previous chain \eqref{eq:Lambda-zero}
is then replaced by a new one~\cite{KlinkhamerVolovik2008-PRD77}
\beq
\Lambda\;\equiv\;\rho_\text{V} \;=\; \widetilde{\epsilon}_\text{V}
\;\stackrel{\text{\textcircled{\sffamily\tiny 1}}}{=}\; -P_\text{V}
\;\stackrel{\text{\textcircled{\sffamily\tiny 2}}}{=}\; P_\text{M}
\;=\;w_\text{M}\,\rho_\text{M} > 0\,,
\label{eq:Lambda-propto-rhoM}
\eeq
for thermal matter with energy density $\rho_\text{M}>0$
and equation-of-state parameter $w_\text{M}> 0$.
A similar connection between the value of the cosmological constant
(vacuum energy density) and that of the averaged energy density of matter
in the Universe
was already found in the very first paper on the topic~\cite{Einstein1917}.
Qualitatively, result \eqref{eq:Lambda-propto-rhoM} points toward a
possible solution of the cosmological constant problems Nos. 2 and 3
from the Introduction.

\textbf{Conclusion 2}:
\emph{The gravitating vacuum energy density of a
perfect (Lorentz-invariant) quantum vacuum in equilibrium vanishes
by the self-tuning of a conserved relativistic variable $q$ and
the gravitating  vacuum energy density of a perturbed imperfect
(Lorentz-noninvariant) quantum vacuum in equilibrium remains small
compared to the microscopic energy density
$\epsilon(q)$, namely, proportional to the Lorentz-violating perturbation.}

\section{COSMOLOGY}
\label{sec:cosmology}

In Sec.~\ref{sec:vacuum-energy}, only the equilibrium situation was considered,
which is static by definition. But what about the vacuum energy density in
Hubble's \emph{expanding} Universe?
This is a difficult question as it concerns the
exchange of energy between the deep vacuum (described in part by the
microscopic variable $q$) and
the low-energy degrees of freedom (described by the standard model
of elementary particles and general relativity).
There is no definite answer yet.

For the moment, two different approaches have been followed:
\begin{enumerate}
\item
to investigate how, starting far away from equilibrium
in a very early phase of the universe, the vacuum
may reach an equilibrium state~\cite{KlinkhamerVolovik2008-PRD78,KlinkhamerVolovik2008-JETPL};
\item
to investigate  how, starting from a postulated equilibrium state at very late times,
a universe can arise which resembles our present Universe~\cite{Klinkhamer2008}.
\end{enumerate}
Here, there is only space to sketch the basic ideas and to give the main results.

\subsection{Vacuum Dynamics}
\label{subsec:vacuum-dynamics}

Vacuum dynamics has been studied with a particular realization of
the $q$ variable. The idea is to start from the four-form field
$F_{\mu\nu\rho\sigma}(x) \equiv \nabla_{[\mu}A_{\nu\rho\sigma]}(x)$,
which is known to have no propagating degrees of freedom in flat
\mbox{4-dimensional} spacetime.
Writing $F_{\mu\nu\rho\sigma}(x) \propto F(x)\,\epsilon_{\mu\nu\rho\sigma}$,
the scalar $F(x)$ has been found to play the role of our $q$ variable.

A second ingredient for the study of nontrivial vacuum dynamics is
the realization that Newton's constant $G_\text{N}$ must be replaced by
a gravitational coupling parameter $G$ which
depends on the state of the vacuum and thus on the vacuum variable $F$.
Such a $G(F)$ dependence can be expected to occur on general grounds.
Moreover, a $G(F)$ dependence allows
the cosmological ``constant'' to change with time, which is otherwise
prohibited by the Bianchi identities and energy-momentum conservation.

With simple \emph{Ans\"{a}tze}
for the microscopic energy density $\epsilon(F)$
and the gravitational coupling parameter $G(F)$,
the field equations can be solved to determine the dynamic behavior
of the vacuum energy density in the context of a flat
Friedmann--Robertson--Walker (FRW) universe. As always, it is useful to
work with dimensionless variables (e.g., a dimensionless comoving time $\tau$),
where the scale is set by the microscopic energy density $\epsilon(q)$
which can be assumed to have a Planckian value.
Recall $E_\text{Planck} \equiv \sqrt{\hbar\, c^5/G_\text{N}} \approx
1.2 \times 10^{28}\;\text{eV}$ and
$t_\text{Planck} \equiv \hbar/E_\text{Planck} \approx
5 \times 10^{-44}\;\text{s}$.

\begin{figure*}[t]
\centering
\includegraphics[width=178mm]{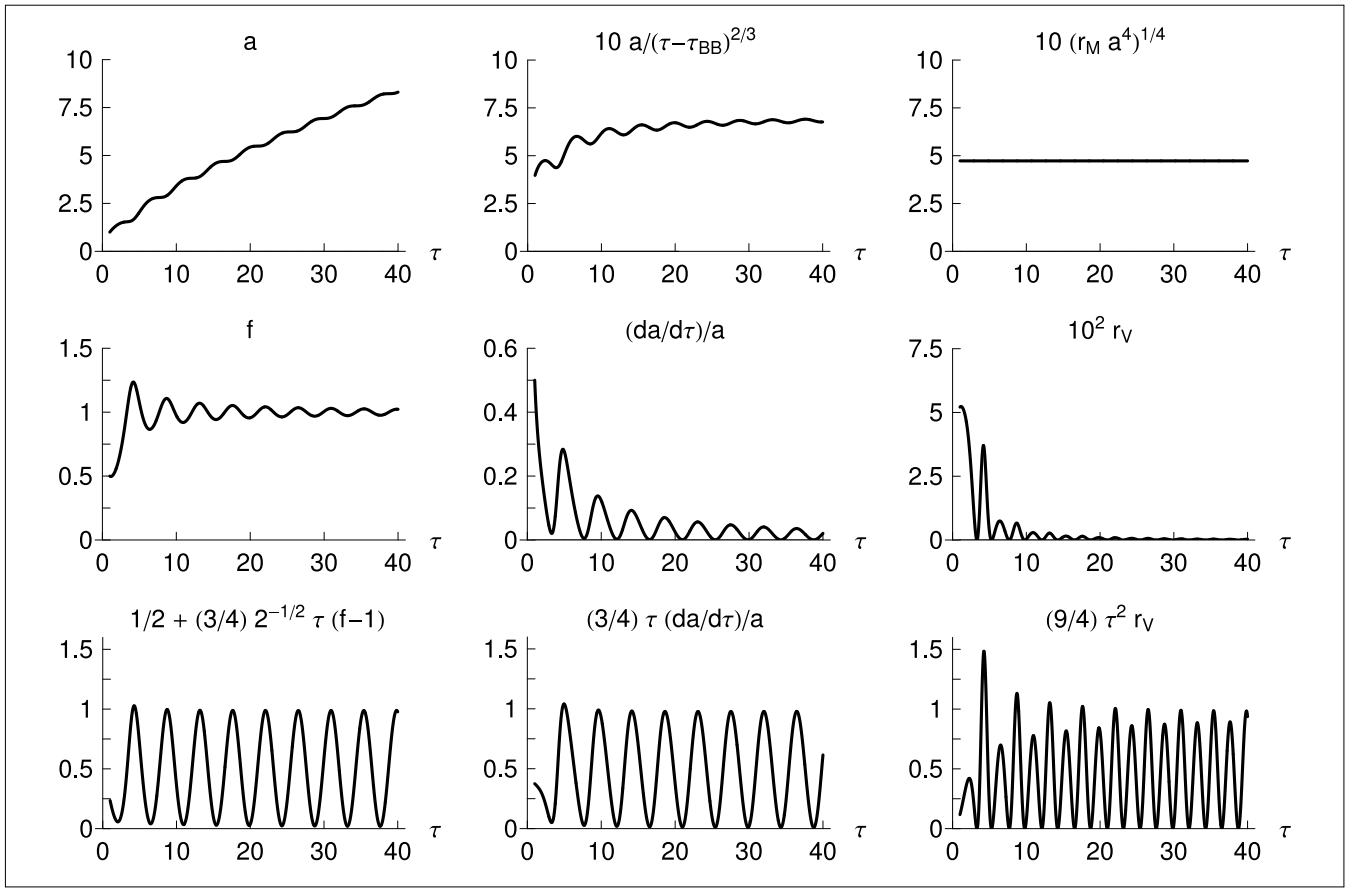}
\caption{Flat FRW model universe~\cite{KlinkhamerVolovik2008-PRD78}
with ultrarelativistic matter
($w_\text{M} \equiv P_\text{M}/\rho_\text{M} =1/3$)
and dynamic vacuum energy
density ($w_\text{V} \equiv P_\text{V}/\rho_\text{V} =-1$),
for initial boundary conditions
$(a,h,f,r_\text{M})=(1,\, 1/2,\, 1/2,\, 1/20)$ at $\tau=1$.
The $q$--type variable $F$ has been made dimensionless and this
dimensionless variable $f$ is seen to approach an equilibrium value
$f=1$ via rapid (Planck-scale) oscillations. Further shown are the dimensionless
Hubble parameter $h(\tau) \equiv (\dd a/\dd \tau)/a$
for scale factor $a(\tau)$ and the
dimensionless energy densities $r_\text{M}(\tau)$ and $r_\text{V}(\tau)$.
The effective parameter $\tau_\text{BB}$ in the middle top-row panel
has been set to the value $-3$.
The three bottom-row panels display the asymptotic behavior $|f-1| \propto 1/\tau$,
$h \propto 1/\tau$, and $r_\text{V} \propto 1/\tau^2$.\vspace*{-100mm}}
\label{fig:1}
\end{figure*}

The results obtained~\cite{KlinkhamerVolovik2008-PRD78} may be relevant
to the early history of the Universe and can be summarized as follows
(see also Fig.~\ref{fig:1} on page 5 of this contribution):
\begin{itemize}
\item
the discovery of a  mechanism of \textbf{vacuum-energy-density decay}, which,
starting from a ``natural'' Planck-scale value at very early times,
leads to the correct order of magnitude  for
the present cosmological constant;
\item
the realization that a substantial part of the \textbf{inferred Cold Dark Matter}
may come from the oscillating part of the vacuum energy density;
\item
the identification of the important role
of \textbf{oscillations of the vacuum variable}
$\boldsymbol{q}$ (with an oscillation period of the order of $t_\text{Planck}$),
which drive the vacuum-energy-density oscillations
responsible for the two first results;
\item
the derivation of an \textbf{$\boldsymbol{f(R)}$ modified-gravity model}
of a particular type~\cite{KlinkhamerVolovik2008-JETPL} that describes
the over-all gravitational effects of the rapidly oscillating vacuum
variable $q$ close to equilibrium.
\end{itemize}

\textbf{Conclusion 3.1}: \emph{The dynamic vacuum variable $q$
allows for the vacuum energy density in a flat FRW universe to relax
from a natural value $\rho_\text{V}\sim (E_\text{Planck})^4$ to
its equilibrium value $\rho_\text{V}=0$, corresponding to Minkowski spacetime.}

\subsection{Closed-Universe Model}
\label{subsec:closed-universe-model}

With the equilibrium physics of the quantum vacuum and the rapid equilibrium
approach reasonably well understood
(cf. Secs.~\ref{sec:vacuum-energy} and \ref{subsec:vacuum-dynamics}),
we now turn to the other side of the coin, namely, the behavior
of the model universe at late times, including the present epoch
(these times are, of course, huge in microscopic units).
Then, the following question arises:
is it possible at all to relate equilibrium boundary
conditions for $\rho_\text{V}(t_\text{eq})$ to
an expanding universe which matches the observations,
even if we are free to choose the type of vacuum-energy dynamics,
$\dd \rho_\text{V}/\dd t\ne 0$?

Inspired by the $q$--theory approach, the following purely phenomenological
\emph{Ansatz} has been proposed~\cite{Klinkhamer2008}
for the time dependence of the vacuum energy density
(and the corresponding energy exchange between vacuum and matter):
\beq
\frac{\dd \rho_\text{V}(t)}{\dd t} = \GammaVM\,\gammaV(t)\;\rho_\text{M}(t)\,,
\label{eq:rhoVdot-Ansatz}
\eeq
with a dimensionless function $\gammaV(t)$ of the comoving time $t$,
normalized by $\gammaV(t_\text{eq})=1$,
and a new fundamental decay constant \mbox{$\GammaVM>0$.}
For pressureless matter, the density $\rho_\text{M}$ in \eqref{eq:rhoVdot-Ansatz}
can be interpreted as corresponding to the cold-dark-matter (CDM) energy density
from observational cosmology, with the baryonic contribution neglected.
Furthermore, the gravitational coupling parameter $G$ is assumed to
equal Newton's constant $G_\text{N}$. In fact,
the model results sketched in the previous subsection give that
$G^{-1}(t) \propto F(t)$
and that fluctuations around $G_\text{N}$ are, at present, suppressed by
a factor $t_\text{Planck}/t_\text{present} \sim 10^{-61}$;
see the two panels with $f(\tau)$ in Fig.~\ref{fig:1} and
Ref.~\cite{KlinkhamerVolovik2008-PRD78} for further details.

\begin{figure*}[t]
\vspace*{2mm}\centering
\includegraphics[width=178mm]{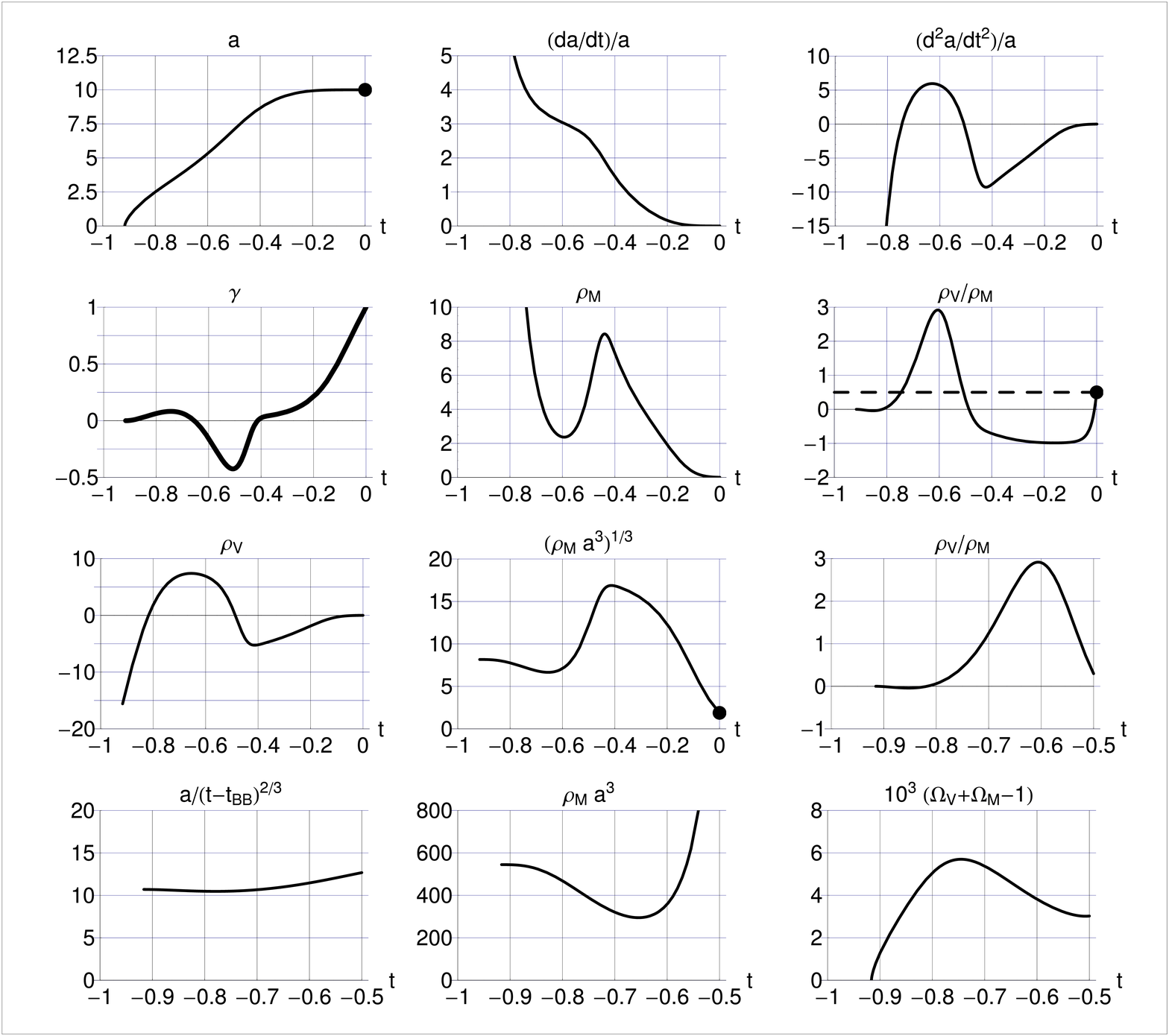}
\caption{Closed FRW model universe~\cite{Klinkhamer2008}
with pressureless matter ($w_\text{M} =0$) and dynamic vacuum energy
density ($w_\text{V}=-1$), for equilibrium boundary conditions
at $t=t_\text{eq}\equiv 0$. The assumed behavior of the vacuum-energy dynamics
is given by \eqref{eq:rhoVdot-Ansatz} with the function $\gamma(t)$
shown in the left-most panel on the second row and
vacuum decay constant $\GammaVM=50$,
in units with $8\pi G_\text{N}/3=c=1$.
The present Universe, with ``cosmic acceleration'' and an energy density
ratio $\rho_\text{V}/\rho_\text{M} \approx 2.75$, would correspond to
this model universe at coordinate time $t=t_0 \approx -0.584$.
The \mbox{``Big Bang''} with 3--sphere radius $a(t_\text{BB})=0$
would occur at coordinate time $t=t_\text{BB}\approx -0.916$.
For the time interval $t \lesssim -0.6$, the model universe has
a nearly standard flat-FRW behavior, as shown by the three bottom-row panels.
For the time interval $-0.6 \lesssim t \lesssim -0.4$,
vacuum energy is transferred to the matter sector, so that
$\rho_\text{M}(t)$ increases with $t$, even though the universe expands.
For the time interval $-0.4 \lesssim t \leq 0$, some energy of the
matter sector is transferred back to the vacuum and the expansion of the
universe slows down to reach the equilibrium radius $a=10$ at $t=0$.
\vspace*{-100mm}}
\label{fig:2}\end{figure*}

With pressureless matter (e.g., CDM) and
vacuum energy density governed by \eqref{eq:rhoVdot-Ansatz},
the idea, now, is to use equilibrium boundary conditions at a coordinate time
$t=t_\text{eq}\equiv 0$ and to solve the Einstein equations of a
closed FRW universe
for time $t$ running in the negative direction
(see Fig.~\ref{fig:2} on page 6 of this contribution).
It indeed turns out to be possible to choose an appropriate function $\gammaV(t)$,
so that a nearly standard Big Bang phase is recovered close to the time
$t=t_\text{BB}<0$ with vanishing 3--volume of the universe,
$V(t_\text{BB})=2 \pi^2\,a(t_\text{BB})^3=0$.
(Note that the boundary conditions have been set at $t=0$, so that
having a Big Bang is a result of the calculation, not an input.)
Moreover, it is possible to identify a moment
$t=t_0$ in the ``life'' of this model universe, which resembles the
presently observed Universe remarkably well:
$(\rho_\text{V0}/\rho_\text{M0})\approx 2.75$,
$(\Omega_{\text{V}0}+\Omega_{\text{M}0})\approx 1.004$,
and $(t_0-t_\text{BB})\approx 14\;\text{Gyr}$.
It is far from trivial that these more or less reasonable values
can be produced at all in our approach.

The main points of this particular toy-model universe~\cite{Klinkhamer2008}
can be summarized as follows:
\begin{itemize}
\vspace*{-0mm}\item
a \textbf{Gibbs--Duhem-type boundary condition} at $t=t_\text{eq}$ with finite
$\rho_\text{V}(t_\text{eq})=(1/2) \; \rho_\text{M}(t_\text{eq})$ for
$w_\text{M}=0$ [this particular value for $\rho_\text{V}(t_\text{eq})$ may result from
the self-tuning of a conserved microscopic variable $q$
to an equilibrium value $q_\text{eq}$ as discussed in Sec.~\ref{sec:vacuum-energy}];
\vspace*{-1mm}\item
a \textbf{Big Bang phase} with $a(t)\propto (t-t_\text{BB})^{2/3}$
for $w_\text{M}=0$ [the power $2/3$ changes to $1/2$ for $w_\text{M}=1/3$];
\vspace*{-1mm}\item
an \textbf{accelerating phase} for ``present times,''
with $\rho_\text{V}/\rho_\text{M}$ of order 1 and an approximately
flat 3--geometry.
\vspace*{2mm}
\end{itemize}

\textbf{Conclusion 3.2}: \emph{An ``existence proof'' has been given
for a model universe with both equilibrium boundary conditions
and a Big Bang.}

\vspace*{-2mm}
\section{OUTLOOK}
\label{sec:outlook}

In the present contribution, we have reviewed one approach to the
statics and dynamics of the quantum vacuum.
This entirely new research topic waits for crucial input from:
\begin{itemize}
\vspace*{-1mm}\item
experiment (e.g., observational cosmology
with a multitude of planned ``dark-energy'' experiments);
\vspace*{-1mm}\item
theory (e.g., the emergent-symmetry approach inspired by condensed-matter physics).
\vspace*{0mm}
\end{itemize}
The coming decennia promise to be exciting for both particle
astrophysics and theoretical physics.

\newpage
\vspace*{0mm}
\newpage
\vspace*{0mm}


\vspace*{-2mm}
\begin{thebibliography}{9}

\bibitem{Einstein1917}
A. Einstein,
Sitzungsber. Preuss. Akad. Wiss., Phys.-Math. Klasse, 1917, p. 142;
translated
in: \emph{The Principle of Relativity},
edited by H.A. Lorentz {\it et al.}
(Dover Publ., New York, USA, 1952), Chap. IX.

\bibitem{Bronstein1933-etal}
(a) M.P. Bronstein,
Phys. Z. Sowjetunion {\bf 3}, 73 \;(1933);
(b) G. Lema\^{i}tre,
Proc. Nat. Acad. Sci. {\bf 20}, 12 \;(1934);
(c) E.B. Gliner,
Sov. Phys. JETP   {\bf 22}, 378 (1966);
(d) Y.B. Zel'dovich,
JETP Lett. {\bf 6}, 316 (1967).


\bibitem{Padmanabhan2003}
T. Padmanabhan,
Phys. Rept.  {\bf 380}, 235 (2003), arXiv:hep-th/0212290.


\bibitem{KlinkhamerVolovik2008-PRD77}
F.R. Klinkhamer and G.E. Volovik,
Phys. Rev. D {\bf 77}, 085015 (2008), arXiv:0711.3170.

\bibitem{KlinkhamerVolovik2008-PRD78}
F.R. Klinkhamer and G.E. Volovik,
Phys. Rev. D {\bf 78}, 063528 (2008),
arXiv:0806.2805.

\bibitem{KlinkhamerVolovik2008-JETPL}
F.R. Klinkhamer and G.E. Volovik,
JETP Lett. {\bf 88}, 289 (2008), arXiv:0807.3896.

\bibitem{Klinkhamer2008}
F.R. Klinkhamer,
Phys. Rev. D {\bf 78}, 083533 (2008), arXiv:0803.0281.

\bibitem{Klinkhamer2008-LV-a-b}
\label{ref:Klinkhamer2008-LV-a-b}
(a) F.R. Klinkhamer,
arXiv:0807.2147;
(b) F.R. Klinkhamer, arXiv:0810.1446.

\bibitem{Volovik2003}
G.E. Volovik,
in: \emph{Patterns of Symmetry Breaking},
edited by H. Arodz {\it et al.}
(Kluwer Academic Publ., Dordrecht, The Netherlands, 2003), p. 381, arXiv:gr-qc/0304061.

\end{thebibliography}
\end{document}